\documentclass[allclo]{FBSart}
\usepackage{epsfig,psfig}

\newcommand{\dens}{{\widehat{w}}}
\newcommand{\wh}{\widehat}
\newcommand{\ul}{\underline}
\newcommand{\ol}{\overline}
\newcommand{\beq}{\begin{equation}}
\newcommand{\eeq}{\end{equation}}
\newcommand{\beqn}{\begin{eqnarray}}
\newcommand{\eeqn}{\end{eqnarray}}

\newcommand\la{\langle}
\newcommand\ra{\rangle}
\newcommand\eps\varepsilon
\newcommand\euler{{\rm e}}
\newcommand\imag{{\imath}}

\newcommand\sgn{{\rm sgn}}

\def\lsim{\mathrel{\rlap{\lower4pt\hbox{\hskip1pt$\sim$}}
    \raise1pt\hbox{$<$}}}
\def\gsim{\mathrel{\rlap{\lower4pt\hbox{\hskip1pt$\sim$}}
    \raise1pt\hbox{$>$}}}

\makeatletter
\def\fmslash{\@ifnextchar[{\fmsl@sh}{\fmsl@sh[0mu]}}
\def\fmsl@sh[#1]#2{%
  \mathchoice
    {\@fmsl@sh\displaystyle{#1}{#2}}%
    {\@fmsl@sh\textstyle{#1}{#2}}%
    {\@fmsl@sh\scriptstyle{#1}{#2}}%
    {\@fmsl@sh\scriptscriptstyle{#1}{#2}}}
\def\@fmsl@sh#1#2#3{\m@th\ooalign{$\hfil#1\mkern#2/\hfil$\crcr$#1#3$}}
\makeatother

\title{Finite-Temperature Field Theory on the Light Front\footnote{Presented by J.R.\ at Light-Cone 2004, Amsterdam, 16 - 20 August, 2004}}
\author{J\"org Raufeisen$^{1}$ and Stanley J. Brodsky$^2$}
\institute{$^1$Institut f\"ur Theoretische Physik der Universit\"at, Philosophenweg 19,\\ 69120 Heidelberg, Germany,
{\tt J.Raufeisen@tphys.uni-heidelberg.de}\\
\smallskip
$^2$Stanford Linear Accelerator Center, Stanford University,\\ 2575 Sand Hill Road, Menlo Park, CA 94025, USA,
{\tt sjbth@slac.stanford.edu}
}
\runningauthor{J.~Raufeisen and S.J.~Brodsky}
\runningtitle{Finite-Temperature Field Theory on the Light-Front}
\begin{document}
\maketitle
\begin{abstract}

The formulation of statistical physics using light-front
quantization, instead of conventional equal-time
boundary conditions, has important advantages for describing relativistic
statistical systems, such as heavy ion collisions.
We develop light-front field theory at finite
temperature and density with special attention to  quantum chromodynamics.
First,
we construct the most general form of the statistical
operator allowed by  the Poincar{\'e} algebra.
In light-front quantization, the
Green's functions of a quark in a medium can be defined in
terms of just
2-component spinors and does not lead to doublers in the transverse directions.  
Since the theory is non-local along the
light cone, we use
causality arguments
to construct a solution to the related zero-mode problem.
A seminal property of light-front Green's functions is that 
they are related to parton densities in coordinate space.
Namely, the diagonal and off-diagonal
parton distributions measured in
hard scattering
experiments can be interpreted as light-front density matrices.
\end{abstract}
%


Dirac's front form of relativistic dynamics \cite{Dirac} has remarkable
advantages in high energy and nuclear physics.
Most appealing is the
simplicity of the vacuum (the ground state of the free theory is also the
ground state of the
full theory) and the existence of boost-invariant light-cone wavefunctions (see
Ref.~\cite{BPPreport} for a review.) This makes light-front quantization a natural candidate for the description of systems for which boost
invariance is an issue,
such as the fireball created in a heavy ion collision or the small-$x$
features  of a nuclear
wavefunction. Until now, however, most applications of Dirac's front form refer to the case of zero temperature.   
It is clearly important to exploit
the advantages of light-front quantization 
also for thermal field theory.
In this talk,
we report our recent findings, see Ref.~\cite{paper}, where we   
applied front-form dynamics to
statistical physics and investigated the prospects and challenges of this
approach for
quantum chromodynamic systems.
Valuable work in
this direction  has already been done by several authors \cite{Beyer, Alves, previous}.


The form of the statistical operator $\dens$ at finite temperature and density can be
obtained from very general considerations.  Our result for $\dens$,
is compatible with the findings of \cite{Beyer,Alves}, {\em
i.e.} $\dens$ is always the exponential of the equal time
energy $\wh P^0$
in the local rest frame of the system.  Our derivation follows  Ref.~\cite{Landau5}.

The light-front Liouville theorem \cite{paper}
($\imag\partial^-\dens=[\wh
P^-,\dens]$) requires that in equilibrium, $\dens$ is a
function of
only those Poincar{\'e} generators which commute with the light-front
Hamiltonian $\wh
P^-$.  In addition, since  systems far apart from each other must be
uncorrelated, the density operator of the combined system has to factorize
into the
density operators of the subsystems. 
Consequently, in equilibrium
$\ln(\dens)$ must
be a linear combination of the additive constants of motion, namely the
four components
of the momentum $\wh P^\mu$ and the $3$-component of the angular momentum vector $\vec J$.
Hence,
\beq\label{eq:general} 
\ln(\dens)=\alpha-\beta\left(u_\nu \widehat
P^\nu-\omega \wh
J_3-\sum_l \mu_l\widehat Q_l\right). 
\eeq  
Here, $\beta$ is the inverse
temperature
and $u_\nu$ is the four velocity of the system, cf.~Ref.~\cite{Alves}.
In addition,
$\omega$ is the angular velocity at which the body rotates.  Additional
conserved
charges $\widehat Q_l$ are included along with their chemical potentials
$\mu_l$.   
In quantum
mechanics, of
course, one can simultaneously specify only charges which commute with each
other, {\em
e.g.}  one cannot specify all
four $P_\nu$
for systems with nonzero angular momentum.

We remark that $T$ is the same as the 
instant-form temperature, but the chemical potential has a different meaning.  
On the light-front, densities are given by
$+$-components of currents, and not by $0$-components.  This is essential
for the proper
generalization of parton distributions (PDFs) to finite temperature, since the
latter are also
defined as $+$-components. Finite temperature PDFs are useful for parton recombination models 
(see {\em e.g.} \cite{Rainer}), even though they cannot be measured in deep inelastic scattering.

We choose $\alpha=0$ as
normalization in Eq.~(\ref{eq:general}), so that the partition function is given by ${\cal Z}={\rm
Tr}\dens$.  
The grand-canonical ensemble can now be written in terms of light-cone wavefunctions $\phi_{n/h}(X)$ as
(let $\vec P_\perp=\vec 0_\perp$),
\beqn\nonumber
\dens&=&\sum_h\sum_{n,n^\prime}\sum_{X,X^\prime}
\exp\left\{-\beta\left[u^+\frac{M_h^2}{2P^+}+\frac{u^-}{2}\sum_ip^+_i-\mu Q\right]\right\}\\
&\times&
\phi_{n/h}(X)\phi^*_{n^\prime/h}(X^\prime)|nX\ra\la n^\prime X^\prime|.
\eeqn
Since the wavefunctions and the masses $M_h$ of the eigenstates can (in principle) be
obtained from discretized light-cone quantization (DLCQ) \cite{DLCQ}, this expression shows
that 
one can also calculate all thermodynamic properties of a field theory from DLCQ.

Furthermore,
since
$\cal Z$ is a Lorentz scalar, all thermodynamic potentials and the entropy
transform as
scalars, {\em e.g.} the Lorentz invariant generalization of the grand-canonical potential
(or of the free energy in the case of $\mu=0$) is
$
\Omega=-T\ln{\cal Z}(V,T,\mu),
$
and the entropy is given by
\beq
S=-\left(\frac{\partial\Omega}{\partial T}\right)_{\mu,V}=-\frac{1}{\cal Z}{\rm Tr}(\dens\ln\dens).
\eeq
The role of the total energy of the system is now played by the expectation value
of $u_\nu\wh P^\nu$,
$
U=\la u_\nu\wh P^\nu\ra.
$
As usual, $\Omega=U-TS-\mu Q$. All known relations between thermodynamic potentials remain valid.

The quantities $\beta$, $u_\nu$, $\omega$  and $\mu_l$, have the meaning of Lagrange multipliers 
that hold the mean values of the constants of motion fixed, while entropy is maximized.
In an ideal gas for example, the maximum entropy is attained for 
occupation numbers given by Fermi-Dirac and Bose-Einstein statistics \cite{Beyer,Alves},
\beq\label{eq:fermibose}
n(u_\nu p^\nu)=\frac{g}{\euler
^{\beta(u_\nu
p^\nu-\mu)}\pm1}
-\frac{g}{\euler^{\beta(u_\nu p^\nu+\mu)}\pm1},
\eeq
assuming that particles carry charge $+1$ and antiparticles charge $-1$.
(Note that $u_\nu p^\nu\ge0$.)
Here, $p^\nu$ is the 4-momentum of a single particle and 
the degeneracy factor for different spin
states is denoted by $g$. 
The Lagrange multipliers define the equilibrium conditions for two systems.
In complete equilibrium with each other, both systems must have the same values of 
temperature, $u_\nu$, $\omega$  and $\mu_l$, {\em i.e.} no internal motion of 
macroscopic parts of the system is possible in equilibrium (at least in the absence of
vortex lines \cite{Landau5}.) 

The simplicity of the light-front vacuum, usually considered an advantage,
seems to bear
problems as far as phase transitions are concerned.  
However, the statistical weight of a
configuration is
maximized for minimal equal-time energy rather than for minimal light-front energy.  
Therefore, the ground
state, {\em i.e.} the state the system is in at $T=0$, is in general
different from the
light-front vacuum. 
For that reason, the authors of Ref.~\cite{Alves} obtain the standard pattern of
spontaneous symmetry breaking in $\phi^4$-theory with negative mass squared. No problem arises from 
$1/k^+$-poles.
In addition, in Ref.~\cite{Beyer} the chiral phase
transition in the
Nambu--Jona-Lasinio Model on the light-front is reproduced.
We conclude that this
approach is poised
for the study of phase transitions in more complicated field theories, such
as QCD.


Until now one could get the impression that thermodynamics and statistical
physics on the
light front are identical to the usual instant form approach, except for a
trivial change
of variables. That this is not the case becomes most clear, when one studies
fermions on
the light-front.
In light-front field theory, the Dirac equations can be written as a set of
two coupled
equations for 2-component spinors, see {\em e.g.} appendix of Ref.~\cite{paper}. Only one
of these
equations contains a time derivative, the other one is a constraint. As a
consequence,
the entire theory can be formulated in terms of 2-component spinors, very
much like a
non-relativistic theory. 

The 
time-ordered Green's functions of a fermion in a medium 
is defined in terms of
the Heisenberg operators of the dynamical field components \cite{Alves,Prem},
\beq\label{eq:green} \imath
G_{\alpha,\beta}(r_1,r_2)=
\la\wh \psi_\alpha(r_1)\wh \psi_\beta^\dagger(r_2)\ra\Theta(r_1^+-r_2^+)
-\la\wh \psi_\beta^\dagger(r_2)\wh \psi_\alpha(r_1)\ra\Theta(r_2^+-r_1^+),
\eeq
where the average
$
\la\dots\ra
$ 
is to be taken with the appropriate ensemble, and $\alpha,\beta\in\{1,2\}$. This definition of the Green's function includes 
the case of zero temperature. 
Therefore, the conventional light-front quantization
at temperature $T=0$ can be formulated in terms of $G_{\alpha,\beta}$ as well.
The Green's function is the fundamental 
object of this approach.
In addition, the retarded ($R$) and
advanced ($A$) Green's functions are defined by the anticommutators
\beqn \imath
G^{R,A}_{\alpha,\beta}(r_1,r_2)&=&\pm
\la\left\{
\wh \psi_\alpha(r_1),\wh \psi_\beta^\dagger(r_2)
\right\}\ra
\Theta(\pm(r_1^+-r_2^+)),
\eeqn 
where the upper sign refers to $G^{R}_{\alpha,\beta}$ and the lower sign to
$G^{A}_{\alpha,\beta}$.
We remark, that
in a gauge theory, it is also necessary to 
include a (path ordered) gauge link along the light-cone by redefining the 
fermion fields, see Ref.~\cite{paper}.

The Green's functions of a fermion in an ideal gas of temperature $T=1/\beta$
were presented first in Ref.~\cite{Alves}. In momentum space, adjusted to  
our notation, they read
\beqn
\widetilde G^{(0)R,A}_{\alpha,\beta}(k)
&=&\delta_{\alpha,\beta} \frac{k^+}{k^2-m^2\pm\imath\epsilon\sgn(uk)}, \\
\widetilde G^{(0)}_{\alpha,\beta}(k)
&=&\delta_{\alpha,\beta} \left({\rm P}
\frac{k^+}{k^2-m^2}
-\imath\sgn(uk)\pi{\rm tanh}\left(\frac{uk}{2T}\right)k^+\delta(k^2-m^2)\right),
\eeqn
where ${\rm P}$ refers to principle value prescription.
The pole prescriptions $\pm\imath\epsilon\sgn(uk)$ for the retarded and advanced Green's functions
are another manifestation of the special meaning of the equal-time energy. These prescriptions
ensure that $G^{(0)R}_{\alpha,\beta}(r)$ vanishes outside the forward lightcone,
while $G^{(0)A}_{\alpha,\beta}(r)$ is non-vanishing only inside the backward lightcone. 
Most importantly, knowledge of the correct pole prescription eliminates ambiguities 
in the definition of the non-local 
operator $1/k^+$, which appears in the free light-cone Hamiltonian. 
However, the correct prescription for the $1/k^+$-pole depends on the 
type of Green's function and on the value of the other momentum components. 

Another remarkable property of the light-front Green's functions is, that
if the theory is discretized on a lattice
in coordinate space,
the factor $k^+$ in the numerator
leads to only one pair of fermion doublers. 
For a transverse lattice lattice approach to finite temperature $SU(\infty)$, see Ref.~\cite{Dalley}.
Moreover,
it is known that no fermion doubling problem occurs in DLCQ and 
one can perform DLCQ calculations for a fixed value of the total charge 
without any sign problem. 

 
In the limit $r^+\to 0^\pm$, the time-ordered Green's function
$G_{\alpha,\beta}$ is closely related to the one-particle density matrices for 
fermions and antifermions,
$q_{\alpha,\beta}(\ul r_1,\ul r_2)$ and $\ol q_{\alpha,\beta}(\ul r_1,\ul r_2)$.
In the 2-component theory, the separation of fermion and antifermion
distributions requires the evaluation of a Fourier-integral
of the Green's function. For quarks, one has
\beq\label{eq:quark}
q_{\alpha,\beta}(k^+,\ul R,\vec r_\perp)
=-\frac{\imath}{4\pi}
\int dr^-\euler^{+\imath k^+r^-/2}G_{\alpha,\beta}(r_2^+\to0^-,\ul r_1,r_2^+,\ul r_2).
\eeq
For antiquarks,
the limit $r^+\to 0$ 
is taken from the other side 
to obtain the correct order of creation and annihilation operators
\cite{paper}.
Since $G_{\alpha,\beta}(r_1,r_2)$ often depends only on the difference
$r=r_1-r_2$, we introduce the variables
$
R=(r_1+r_2)/2$ and
$r=r_1-r_2$.
The density matrix
$q_{\alpha,\beta}$ is related 
to the so-called Wigner function by a
Fourier transform over $r_\perp$. 
We remark that all properties of the quantum mechanical density matrix, such as 
hermiticity and
positivity of the diagonal matrix elements, also apply to the light-front density 
matrix in $A^+=0$ gauge, since this gauge has only states with positive norm and
no unphysical degrees of freedom.
However, the  light-front density matrix has matrix elements 
that are off-diagonal in Fock-space.
The object defined in Eq.~(\ref{eq:quark}) is similar to the Wigner function 
introduced in Ref.~\cite{wigner}. 

The fermion density matrix contains all information about single-quark
properties. It depends on 6 variables and is a $2\times2$ matrix in spinor
space, which can be written as a linear combination of Pauli spin matrices.
The coefficients are the density matrices for
unpolarized, longitudinal, and
transverse  spin distributions.
The diagonal matrix elements in coordinate space
of $q_{\alpha,\beta}(k^+,\ul R,\vec r_\perp)$, {\em i.e.}
the ones with $\vec r_\perp=\vec 0_\perp$, are closely related to 
the usual PDFs. For instance,
the unpolarized collinear quark density is given by
\beq
q(k^+)=\frac{1}{2}\int d^3R\delta_{\alpha,\beta}q_{\beta,\alpha}(k^+,\ul R,\vec r_\perp=\vec 0_\perp).
\eeq
This parton density is normalized 
such that $\int_0^\infty dk^+q(k^+)=q$, the total 
number of quarks in the system. 

The off-diagonal matrix elements 
of
$q_{\alpha,\beta}$
are related to generalized parton distributions
(GPDs) \cite{gpdf} by Fourier transform \cite{paper}.
The precise relation depends on the kinematics, and one has to
distinguish four different domains in deeply virtual Compton scattering (DVCS). In particular,
for skewedness $\zeta=0$, GPDs can be identified as 
impact parameter dependent parton densities \cite{Matthias}. In the case of no helicity flip we find,
\beqn
q(k^+,\vec b)
&=&\int dR^-\frac{1}{2}\delta_{\beta,\alpha}
q_{\alpha,\beta}(k^+,R^-,\vec b,\vec r_\perp=\vec 0_\perp)\\
&=&
\frac{1}{4\pi k^+}\left<\sum_{\lambda}
\wh b^\dagger(k^+,\vec b,\lambda)\wh b(k^+,\vec b,\lambda)
\right>,\\
q(k^+,\vec b)dk^+
&=&\int \frac{d^2\Delta_\perp}{(2\pi)^2}\euler^{\imath\vec b\cdot\vec\Delta_\perp}H(X,\zeta=0,t),
\eeqn
We use the notation of Radyushkin here \cite{gpdf}.
The creation operator of a quark at impact parameter $\vec b$
is given by
\beq
\wh b(k^+,\vec b,\lambda)=\int\frac{d^2k_{\perp}}{(2\pi)^2}\euler^{-\imath\vec k_{\perp}\cdot\vec b}\wh b(k^+,\vec k_{\perp},\lambda).
\eeq
The destruction operator $\wh b^\dagger(k^+,\vec b,\lambda)$ is defined analogously.
We stress that for $\zeta\neq 0$,
GPDs are in general not 
probability distributions but density matrices, which do not need to be positive.

The light-front density matrix is a natural extension of
the parton model to quantum mechanics:
classical parton densities are replaced by a density matrix.
It would be interesting to identify other hard processes besides DVCS
which are sensitive
to the quantum mechanical nature of parton distributions.
Most important however is the connection between the density matrix and
the fermion Green's function, because that establishes a common language 
for high energy scattering and statistical QCD.

In summary,
we have presented a new formalism for  analyzing relativistic statistical systems based on light-front quantization. The new formalism provides a boost-invariant generalization of thermodynamics, and thus it
has direct applicability to the QCD analysis of heavy ion collisions and other systems of relativistic particles.

\medskip
\noindent {\bf Acknowledgments}: We  thank the organizers of the
Light-Cone 2004 workshop in Amsterdam for inviting us to this stimulating meeting.
This work was supported by the
Feodor Lynen
Program of the Alexander von Humboldt Foundation
and by the U.S.\ Department
of Energy at
SLAC under Contract No.~DE-AC02-76SF00515.

\end{document}